\documentclass[epj]{webofc}
\usepackage[utf8]{inputenc}
\usepackage[varg]{txfonts}   
\usepackage{booktabs}
\usepackage{xcolor}
\definecolor{darkred}{rgb}{0.4,0.0,0.0}
\definecolor{darkgreen}{rgb}{0.0,0.4,0.0}
\definecolor{darkblue}{rgb}{0.0,0.0,0.4}
\usepackage[bookmarks,linktocpage,colorlinks,
    linkcolor = darkred,
    urlcolor  = darkblue,
    citecolor = darkgreen]{hyperref}
\usepackage{subfigure}

\wocname{EPJ Web of Conferences}
\woctitle{Lattice2017}

\graphicspath{{plots/}}
\newcommand{\g}[1]{\gamma_{#1}} 
\renewcommand{\l}{\left}
\renewcommand{\r}{\right}

\newcommand{\diag}{\mathrm{diag}}  

\newcommand{\gev}{\,\mathrm{GeV}}

\newcommand{\mev}{\,\mathrm{MeV}}
\newcommand{\fm}{\,\mathrm{fm}}

\newcommand{\MSbar}{\overline{\mathrm{MS}}}

\newcommand{\tsep}{t_\mathrm{sep}}

\newcommand{\Mpi}{M_\pi}

\newcommand{\avgx}[2]{\langle x \rangle_{#2 u #1 #2 d}}

\begin{document}

\selectlanguage{english}

\title{
Nucleon average quark momentum fraction with $N_\mathrm{f}=2+1$ Wilson fermions
}

\author{
\firstname{Konstantin} \lastname{Ottnad}     \inst{1}   \fnsep \thanks{Speaker, \email{kottnad@uni-mainz.de}} \and
\firstname{Tim}        \lastname{Harris}     \inst{2}   \and
\firstname{Harvey}     \lastname{Meyer}      \inst{1,2} \and
\firstname{Georg}      \lastname{von Hippel} \inst{1}   \and
\firstname{Jonas}      \lastname{Wilhelm}    \inst{1}   \and
\firstname{Hartmut}    \lastname{Wittig}     \inst{1,2}
}

\institute{
PRISMA Cluster of Excellence and Institut f\"ur Kernphysik, Johann-Joachim-Becher-Weg 45, University of Mainz, 55099 Mainz, Germany
\and
Helmholtz Institute Mainz, University of Mainz, 55099 Mainz, Germany
}

\abstract{
We report on an analysis of the average quark momentum fraction of the nucleon and related quantities using $N_\mathrm{f}=2+1$ Wilson fermions. Computations are performed on four CLS ensembles covering three values of the lattice spacing at pion masses down to $M_\pi \approx 200\,\mathrm{MeV}$. Several source-sink separations ($\sim 1.0\,\mathrm{fm}$ to $\sim 1.4\,\mathrm{fm}$) are used to assess the excited-state contamination. To gain further insight, the generalized pencil-of-functions approach has been implemented to reduce the excited-state contamination in the relevant two- and three-point functions. Preliminary results are shown for the isovector nucleon charges from vector, axial vector and tensor derivative (twist-2) operators.
}

\maketitle

\section{Introduction}\label{sec:introduction}
In this proceedings contribution we present a nucleon structure calculation by the Mainz group concerning twist-2 operator insertions with focus on the nucleon quark momentum fraction. It is complementing a similar study of local charges and electromagnetic form factors which has also been presented at this conference \cite{thiscontrib181}. \par

The nucleon quark momentum fraction is defined as the first momentum of the distribution of unpolarized quarks
\begin{equation}
 \langle x\rangle_{q} = \int_0^1 dx \, x \cdot \l[q(x) + \bar{q}(x)\r] \,.
 \label{eq:def_average_x}
\end{equation}
Similarly, one defines the first moment helicity and transversity moments $\langle x \rangle_{\Delta q}$ and $\langle x \rangle_{\delta q}$ from distributions of correspondingly polarized quarks $\Delta q$ and $\delta q$. In lattice QCD, suitable twist-2 operator insertions for nucleon three-point functions are required to compute these observables 
\begin{align}
 \mathcal{O}^{vD}_{\mu\nu}     &= \bar{q} \gamma_{\l\{ \mu \r.} \stackrel{\leftrightarrow}{D}_{\l.\nu\r\}} q \,, \label{eq:vD} \\
 \mathcal{O}^{aD}_{\mu\nu}     &= \bar{q} \gamma_{\l\{ \mu \r.} \gamma_5 \stackrel{\leftrightarrow}{D}_{\l.\nu\r\}} q \,, \label{eq:aD} \\
 \mathcal{O}^{tD}_{\mu\nu\rho} &= \bar{q} \sigma_{[\mu\l\{\nu\r.]} \stackrel{\leftrightarrow}{D}_{\l.\rho\r\}} q \,, \label{eq:tD}
\end{align}
where $\{...\}$ denotes symmetrization over indices with subtraction of the trace and $[...]$ denotes anti-symmetrization. The symmetric derivative is defined as $\stackrel{\leftrightarrow}{D}_\mu = \frac{1}{2}\bigl( \stackrel{\rightarrow}{D}_\mu - \stackrel{\leftarrow}{D}_\mu \bigr)$. \par

\section{Form factor decomposition}\label{sec:FF_decomposition}
For the vector-derivative insertion $\mathcal{O}^{vD}_{\mu\nu}$, the form factor decomposition of the corresponding nucleon matrix element reads
\begin{align}
  \l\langle N(p_f,s_f) \l| \mathcal{O}^{vD}_{\mu\nu} \r| N(p_i,s_i)\r\rangle = \bar{u}(p_f,s_f) & \left[ \gamma_{\left\{\mu\right.} \bar{P}_{\left.\nu\right\}} A_{20}(Q^2) - \frac{\sigma_{\left\{ \mu\alpha\right. } Q_\alpha Q_{\left.\nu\right\}} }{2m_N} B_{20}(Q^2) \right. \notag \\
  & \ \left. + \frac{Q_{\left\{\mu \right.}Q_{\left.\nu\right\}}}{m_N}  C_{20}(Q^2)  \right] u(p_i,s_i) \,,
  \label{eq:matrix_element}
\end{align}
where $u(p_i,s_i)$, $\bar{u}(p_f,s_f)$ are Dirac spinors with initial (final) state momentum $p_i$ ($p_f$) and spin $s_i$ ($s_f$), $m_N$ the nucleon mass, $\g{\mu}$ a Dirac matrix and $\sigma_{\mu\nu} = \frac{1}{2}\bigl[\g{\mu},\g{\nu}\bigr]$. We work in Euclidean spacetime, e.g. $Q$ denotes Euclidean four-momentum transfer with $Q^2=-q^2$, $q=p_f-p_i$. For the labeling of the generalized form factors $A_{20}(Q^2)$, $B_{20}(Q^2)$ and $C_{20}(Q^2)$ on the r.h.s. and more details on the form factor decomposition for generalized parton distributions we refer to Ref.~\cite{Hagler:2004yt}. Spin-projecting with $\Gamma_0=\frac{1}{2}(1+\gamma_0)$ and $\Gamma_z = \Gamma_0 (1+i\gamma_5\gamma_3)$ and considering zero momentum transfer, we have to compute the ratio
\begin{equation}
 R^{vD}_{\mu\nu}(t_f,t,t_i)=\frac{C_{\mathrm{3pt}}^{\mu\nu}(\vec{q}=0, t_f, t, t_i; \Gamma_z)}{C_\mathrm{2pt}(\vec{q}=0,t_f-t_i;\Gamma_0)} \rightarrow \left\{
 \begin{array}{ll} 
  -\frac{3}{4} m \avgx{\pm}{} & \mbox{for} \ \mu=0 \\[0.5em] 
  +\frac{1}{4} m \avgx{\pm}{} & \mbox{for} \ \mu=1,2,3 
 \end{array} \right. \,,
 \label{eq:ratio_vD}
\end{equation}
for $t_f-t\gg1$, $t-t_i\gg1$, where $t\equiv t_\mathrm{ins}$ denotes the time of the operator insertion. Subscripts $f$ and $i$ again denote final and initial state, respectively. In the above expression we identified $\left<x\right>_{u\pm d} \equiv A_{20}(0)$, where the subscript notation $u\pm d$ refers to choosing either isoscalar (``$+$'') or isovector (``$-$'') combinations in the operator insertion. For $\mathcal{O}^{aD}_{\mu\nu}$ and $\mathcal{O}^{tD}_{\mu\nu\rho}$ one finds similar relations from the corresponding form factor decompositions
\begin{align}
 R^{aD}_{\mu\nu}(t_f,t,t_i) \rightarrow& -\frac{i}{2} m \avgx{\pm}{\Delta} && \mbox{for} \ \mu=3, \ \nu=0 \,, \label{eq:ratio_aD} \\
 R^{tD}_{\mu\nu\rho}(t_f,t,t_i) \rightarrow& +\frac{i}{4} m \left( 2\delta_{0\rho} - \delta_{0\mu} - \delta_{0\nu}\right) \avgx{\pm}{\delta} && \mbox{for} \ \mu=0, \ \nu=1, \ \rho=2 \,. \label{eq:ratio_tD}
\end{align}

\section{Computation of two-point and three-point functions}\label{sec:computation}
For the computation of the relevant quark-connected two- and three-point functions in Eq.~(\ref{eq:ratio_vD}) we employ the truncated solver method \cite{Bali:2009hu,Shintani:2014vja}. It relies on performing a (large) number $N_{LP}$ of cheap, low-precision inversions and using a small number of high-precision measurements $N_{HP}$ to correct for the resulting bias in the desired expectation value
\begin{equation}
 \langle\mathcal{O}\rangle = \langle\frac{1}{N_{LP}}\sum_{i=1}^{N_{LP}}\mathcal{O}_i^{LP}\rangle + \langle \mathcal{O}_\mathrm{bias}\rangle \,, \quad \mathcal{O}_\mathrm{bias} = \frac{1}{N_{HP}}\sum_{i=1}^{N_{HP}}(\mathcal{O}_i^{HP} - \mathcal{O}_i^{LP}) \,.
 \label{eq:truncated_solver_method}
\end{equation}
Typically we use between 16 and 48 low-precision measurements and a single, high-precision measurement for the bias correction. Depending on the observable and the ensemble this results in a gain of a factor $\sim 2 $ to $3$ in computer time. For the computation of three-point functions we use sequential inversions through the sink, with the final state produced at rest, i.e. $\vec{p}_f=0$. We remark that in this study we focus on the computation of isovector observables, for which quark-disconnected contributions cancel. \par

\section{Ensembles}\label{sec:ensembles}
In this study we use gauge configurations generated by CLS with $N_\mathrm{f}=2+1$ dynamical flavors of non-perturbatively Clover improved Wilson quarks \cite{Bruno:2014jqa}. At the time of the conference we had analyzed data on four ensembles that are listed in Tab.~\ref{tab:ensembles}. While this is only a subset of the ensembles employed in the calculation of electromagnetic form factors in Ref.~\cite{thiscontrib181}, they still cover three values of the lattice spacing $a$ between $\sim 0.05\fm$ and $\sim 0.09 \fm$. Pion masses vary between $200\mev$ to $280\mev$ \cite{Bruno:2016plf}. We plan to extend our analysis by additional ensembles, including a fourth, intermediate value of the lattice spacing corresponding to $\beta=3.46$, as well as more pion masses. This should ultimately enable us to perform a reliable chiral and continuum extrapolation for all observables. \par 

\begin{table}[thb]
\centering
 \begin{tabular*}{.92\textwidth}{@{\extracolsep{\fill}}lccccclr}
  \hline\hline
  ID & $\beta$ & $a/\fm$ & $L/a$ & $M_\pi/\mev$ & $M_\pi L$ & $\tsep/\fm$ & $N_\mathrm{meas}$ \\
  \hline\hline
   H105 & 3.40 & 0.087 & 32 & 280 & 3.9 & 1.0, 1.2, 1.4      & 48912 \\
   N200 & 3.55 & 0.064 & 48 & 280 & 4.4 & 1.0, 1.2, 1.3, 1.4 & 20364 \\
   D200 & 3.55 & 0.064 & 64 & 200 & 4.2 & 1.0, 1.2, 1.3, 1.4 & 32672 \\
   J303 & 3.70 & 0.050 & 64 & 260 & 4.1 & 1.0, 1.1, 1.2, 1.3 &  5856 \\
  \hline\hline
  \vspace*{0.1cm}
 \end{tabular*}
 \caption{The CLS ensembles used in this study. In addition to $\beta$, the lattice spacing $a$, $L/a$, $M_\pi$ and $M_\pi L$, we have included the source-sink separations $\tsep$ in physical units and the number of measurements $N_\mathrm{meas}$.}
 \label{tab:ensembles}
\end{table}

\section{Renormalization}\label{sec:renormalization}
We have performed the non-perturbative renormalization using the Rome-Southampton method \cite{Martinelli:1994ty} for the two lower values of $\beta$. Tab.~\ref{tab:renormalization} contains our results in the $\MSbar$ scheme at a scale of $\mu=2\,\gev$ for both irreducible representations for each of the three operators in Eqs.~(\ref{eq:vD}),~(\ref{eq:aD})~and~(\ref{eq:tD}). Since the generation of ensembles with periodic boundary conditions is not feasible at $\beta=3.7$ due to the expected freezing of the topological charge, we could not obtain non-perturbative measurements at the finest lattice spacing. Therefore, we have to resort to a linear extrapolation for this value of $\beta$. We have scaled the corresponding statistical errors by a factor of 10 to account for the systematic uncertainty of this procedure. In order to remedy this situation in the future, and because we plan to include an ensemble at $\beta=3.46$, we will extend the computation of renormalization factors to this intermediate value of $\beta$. This should yield a more reliable extrapolation for the renormalization factors at $\beta=3.7$. \par
\begin{table}[thb]
 \centering
  \begin{tabular}{@{\extracolsep{\fill}}lcccccc}
   \hline\hline
   $\beta$ & $Z^{\MSbar}_{v2a}$ & $Z^{\MSbar}_{v2b}$ & $Z^{\MSbar}_{r2a}$ & $Z^{\MSbar}_{r2b}$ & $Z^{\MSbar}_{h1a}$ & $Z^{\MSbar}_{h1b}$ \\
   \hline\hline
   3.40 & 1.0885(01) & 1.0684(01) & 1.1118(01) & 1.0561(01) & 1.0996(01) & 1.1156(01) \\
   3.55 & 1.1388(01) & 1.1237(01) & 1.1601(01) & 1.1130(01) & 1.1612(01) & 1.1756(01) \\
   3.70 & 1.1850(11) & 1.1745(11) & 1.2045(11) & 1.1653(11) & 1.2178(11) & 1.2307(11) \\
   \hline\hline
   \vspace*{0.1cm}
  \end{tabular}
  \caption{Non-perturbative values for renormalization constants at $\beta=3.40,3.55$. Values for $\beta=3.7$ have been obtained from a linear extrapolation. Results are included for both irreducible representations of all three operators.}
  \label{tab:renormalization}
 \end{table}

\section{Excited states and generalized pencil-of-functions approach} \label{sec:excited_states}
Nucleon structure calculations are known to be hampered by excited-state contamination. This is due to a signal-to-noise problem that prevents one from reaching large source-sink separations for nucleon three-point functions as would be required for e.g. the ratio in Eq.~(\ref{eq:ratio_vD}) to reliably represent the plateau value. A commonly used approach to deal with this issue is the summation method \cite{Maiani:1987by,Capitani:2012gj}. It relates the sum over timeslices of a ratio $R^X(t_f,t,t_i)$, $X=vD,aD,tD$ at a given value of $\tsep$ to the desired ground-state matrix element $\mathcal{M}_0$,
\begin{equation}
 \sum\limits_{t=t_i+2}^{t_f-2} R(t_f,t,t_i) = \mathrm{const} + t_f \cdot \mathcal{M}_0 + \mathcal{O}\left(e^{-\Delta E(t_f-t_i)}\right) \,,
 \label{eq:summation_method}
\end{equation}
where two timeslices at source and sink have to be skipped in the summation for twist-2 operator insertions. The matrix element $\mathcal{M}_0$ can be extracted from a linear fit to the data on the r.h.s for multiple source-sink separations. The leading correction $\sim \exp(- \Delta E (t_f-t_i))$ is more strongly suppressed than the corresponding correction for the plateaux method. However, in practice the efficacy of the summation method is driven by the values of $\tsep$ for which a good statistical precision can be achieved. The resulting statistical error is typically larger than for the plateau method. \par

\begin{figure}[thb]
 \centering
 \subfigure[Eigenvalues]{\includegraphics[totalheight=0.275\textheight]{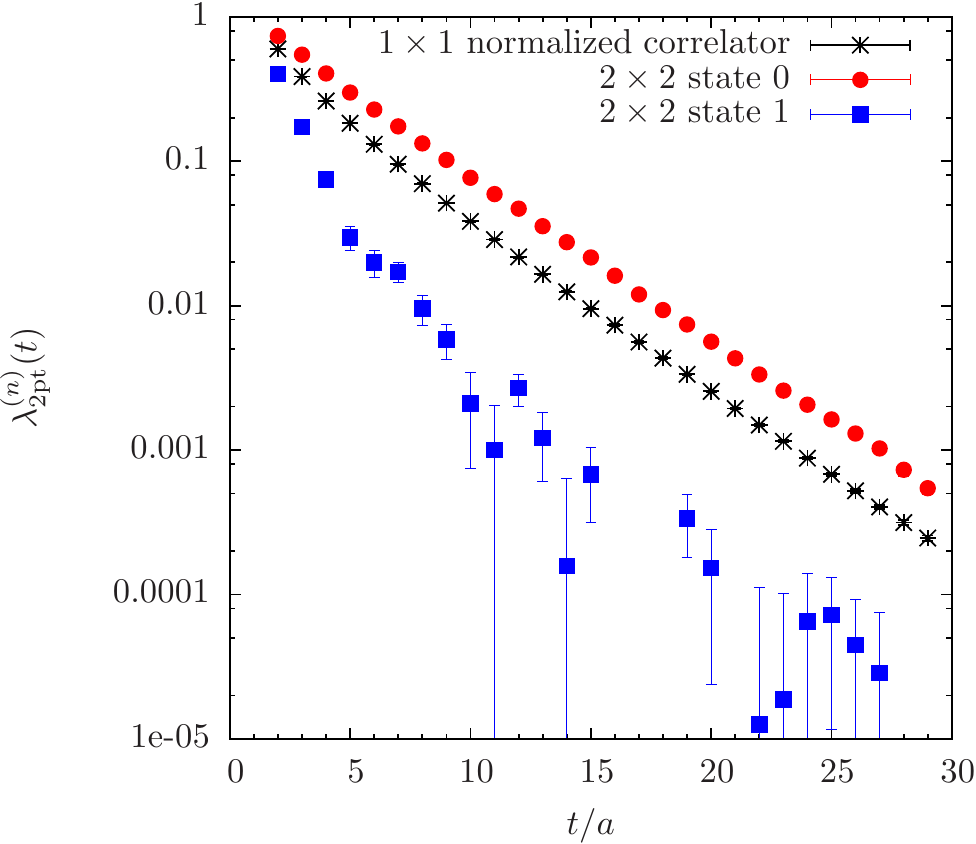}}\quad
 \subfigure[Corresponding eigenvectors]{\includegraphics[totalheight=0.275\textheight]{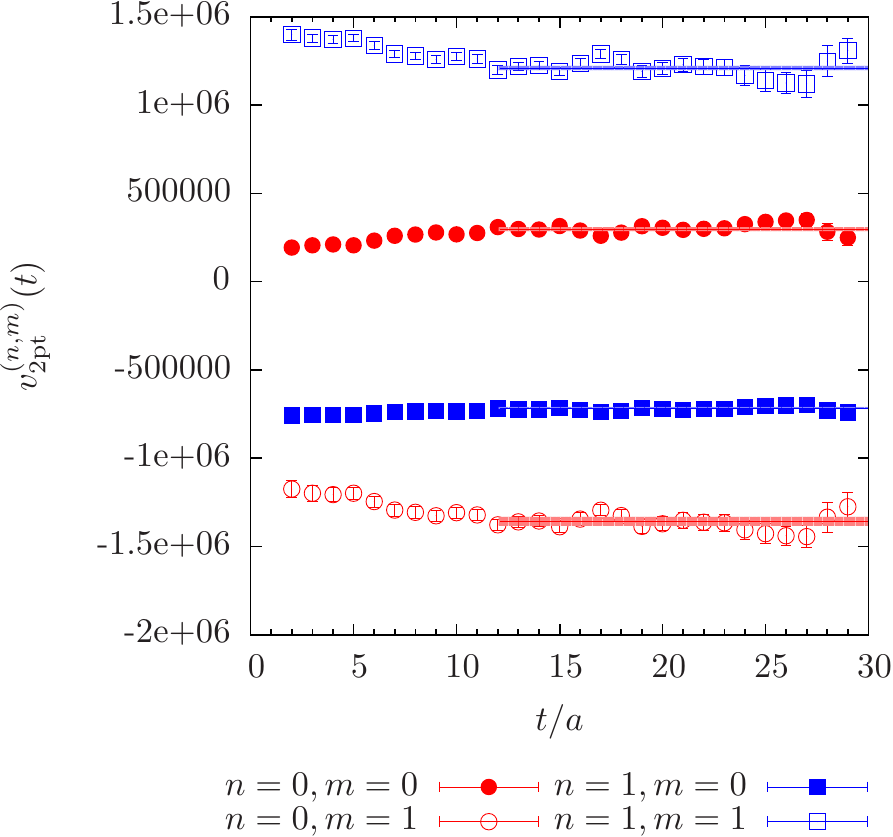}}
 \caption{Principal correlators and eigenvectors for $2\times2$ generalized pencil-of-functions on ensemble D200.}
 \label{fig:GPOF}
\end{figure}

Another way to tackle the excited-state problem is the so-called generalized pencil-of-function (GPOF) approach, which was first applied for baryon calculations in a study of the electromagnetic form factor of the $\Delta$ \cite{Aubin:2010jc,Aubin:2011zz}. The method relies on the fact that an existing operator basis (here just a single operator for the nucleon) can be increased by performing a time-shift
\begin{equation}
 \mathcal{O}_{\Delta t}(t) = \mathcal{O}(t+\Delta t) = \exp(H \Delta t) \, \mathcal{O}(t) \, \exp(-H \Delta t) \,,
 \label{eq:GPOF_operator}
\end{equation}
which yields additional, linearly independent operators. This allows one to build a $(n+1) \times (n+1)$ correlation function matrix of two-point functions for fixed $\Delta t$, $t=t_f-t_i$:
 \begin{equation}
  \mathcal{C}_\mathrm{2pt}(t) = \left(\begin{smallmatrix}
   \langle\mathcal{O}_{0\cdot\Delta t}(t_f) \mathcal{O}^\dag(t_i)\rangle & \dots & \langle\mathcal{O}_{0\cdot\Delta t}(t_f) \mathcal{O}^\dag_{n\cdot\Delta t}(t_i)\rangle \\
   \vdots &  \ddots & \vdots \\
   \langle\mathcal{O}_{n\cdot\Delta t}(t_f) \mathcal{O}^\dag(t_i)\rangle & \dots & \langle\mathcal{O}_{n\cdot\Delta t}(t_f) \mathcal{O}^\dag_{n\cdot\Delta t}(t_i)\rangle \\
  \end{smallmatrix}\right) = \left(\begin{smallmatrix}
   C_\mathrm{2pt}(t) & \dots & C_\mathrm{2pt}(t+n\cdot\Delta t) \\
   \vdots & \ddots & \vdots \\
   C_\mathrm{2pt}(t+n\cdot\Delta t) & \dots & C_\mathrm{2pt}(t+2n\cdot\Delta t)
  \end{smallmatrix}\right) \,.
  \label{eq:GPOF_2pt}
 \end{equation}
For this matrix one can solve a generalized eigenvalue problem in the standard way, which results in eigenvalues $\lambda^{(n)}(t,t_0)$ and a matrix of eigenvectors $V= (\vec{v}^{(0)}(t,t_0), ... ,\vec{v}^{(n)}(t,t_0))$. The results from this procedure are shown for one ensemble in Fig.~\ref{fig:GPOF}. Clearly, the excited-state contamination is reduced in the ground-state principal correlators in the left panel.

Similarly one obtains a (non-symmetric) matrix of three-point functions 
\begin{equation}
 \mathcal{C}_\mathrm{3pt}(t_f,t,t_i) = \left(\begin{smallmatrix}
  C_\mathrm{3pt}(t_f,t,t_i) & \dots & C_\mathrm{3pt}(t_f+n\cdot\Delta,t+n\cdot\Delta t,t_i) \\
  \vdots & \ddots & \vdots \\
  C_\mathrm{3pt}(t_f+n\cdot\Delta t,t,t_i) & \dots & C_\mathrm{3pt}(t_f+2n\cdot\Delta t,t+n\cdot\Delta t,t_i) \,.
 \end{smallmatrix}\right)\,,
 \label{eq:GPOF_3pt}
\end{equation}
which can be diagonalized using the eigenvectors from the two-point case
 \begin{equation}
  \mathcal{C}_\mathrm{3pt}(t_f,t,t_i) \rightarrow V^T \mathcal{C}_\mathrm{3pt}(t_f,t,t_i) V = \Lambda_\mathrm{3pt}(t_f,t,t_i) = \diag(\Lambda^{(0)}, ... ,\Lambda^{(n)})(t_f,t,t_i) \,.
  \label{eq:GPOF_diagonlization}
\end{equation}
To this end, a plateau fit is applied to the eigenvectors to remove the $t$-dependence, as shown in the right panel of Fig.~\ref{fig:GPOF}. Finally, the ratios in Eqs.~(\ref{eq:ratio_vD}),~(\ref{eq:ratio_aD})~and~(\ref{eq:ratio_tD}) are replaced by the new, optimized ratio built from ground-state two- and three-point functions $\Lambda_\mathrm{3pt}^{(0)}(\vec{q}=0,t_f, t_i, t; \Gamma_z)$ and $\lambda_\mathrm{2pt}^{(0)}(\vec{q}=0,t_f-t_i;\Gamma_0)$
\begin{equation}
 \frac{C_\mathrm{3pt}(\vec{q}=0,t_f, t, t_i; \Gamma_z)}{C_\mathrm{2pt}(\vec{q}=0,t_f-t_i;\Gamma_0)} \rightarrow \frac{\Lambda_\mathrm{3pt}^{(0)}(\vec{q}=0,t_f, t, t_i; \Gamma_z)}{\lambda_\mathrm{2pt}^{(0)}(\vec{q}=0,t_f-t_i;\Gamma_0)} \,.
 \label{eq:GPOF_ratio}
\end{equation}
Fitting to this ratio should reduce the excited-state contamination in the final result. Moreover, it typically yields smaller statistical errors than the summation method. In practice, certain restrictions apply, i.e. the possible choices of $\Delta t$ are restricted by the available values of $\tsep$. Currently, the only possible choice is $\Delta t=2$ and building a $2\times 2$ matrix, as can be inferred from the values of $\tsep$ listed in Tab.~\ref{tab:ensembles}. \par

\section{Results and outlook}\label{sec:results}
\begin{figure}[thb]
 \centering
 \subfigure[Lattice and GPOF data for $\avgx{-}{}$ on D200]{\includegraphics[totalheight=0.275\textheight]{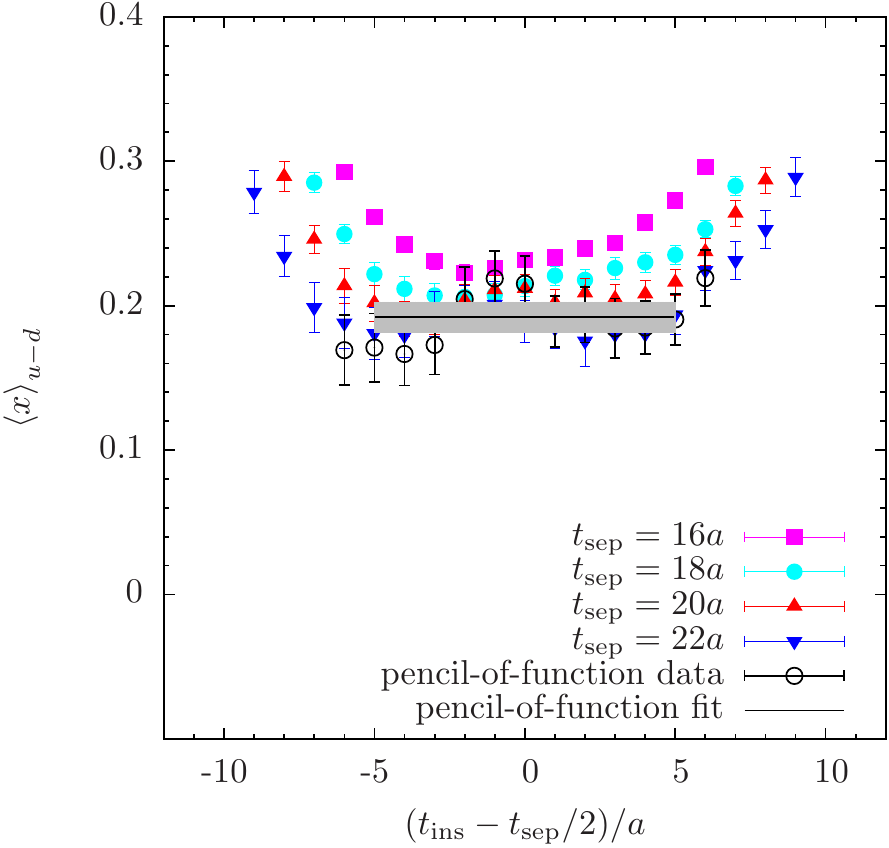}}\quad
 \subfigure[Results for $\avgx{-}{}$ on D200 for different methods.]{\includegraphics[totalheight=0.275\textheight]{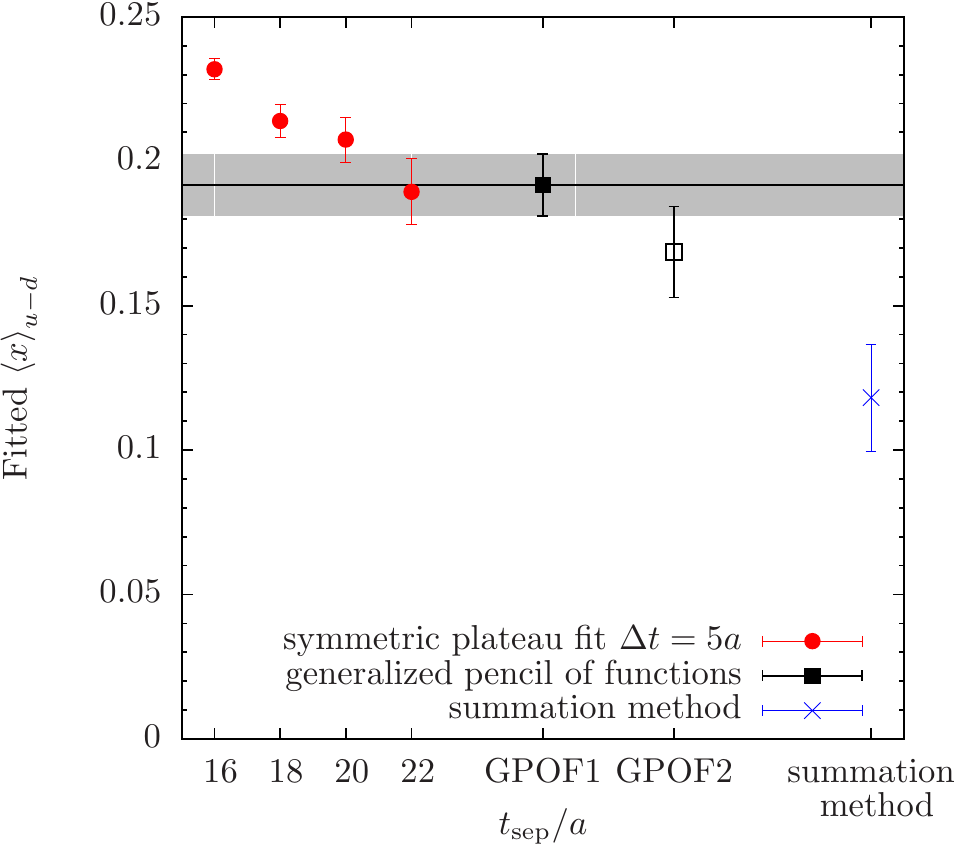}}\\
 \subfigure[Lattice and GPOF data for $\avgx{-}{\Delta}$ on N200]{\includegraphics[totalheight=0.275\textheight]{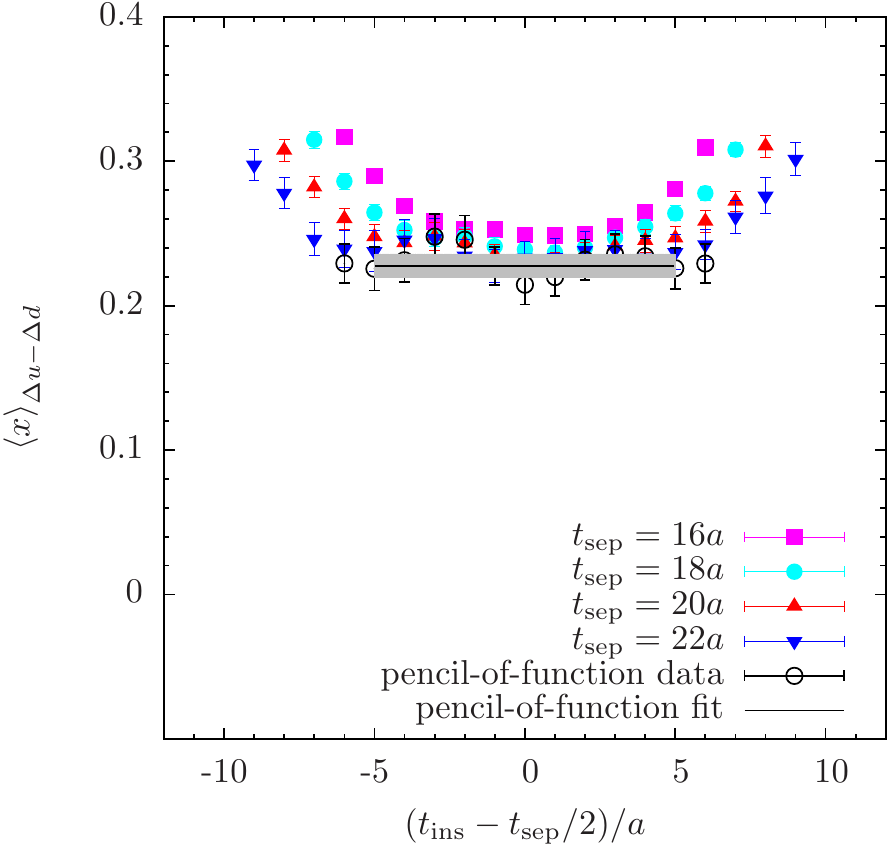}}\quad
 \subfigure[Results for $\avgx{-}{\Delta}$ on N200 for different methods.]{\includegraphics[totalheight=0.275\textheight]{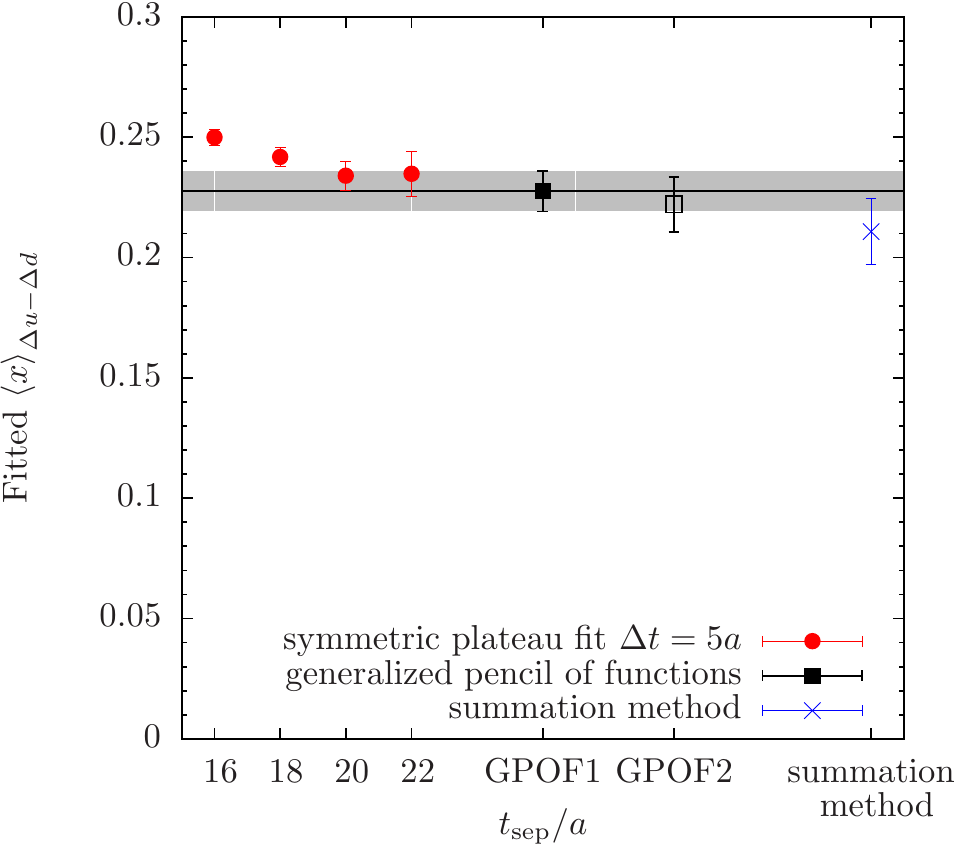}}\\
 \subfigure[Lattice and GPOF data for $\avgx{-}{\delta}$ on J303]{\includegraphics[totalheight=0.275\textheight]{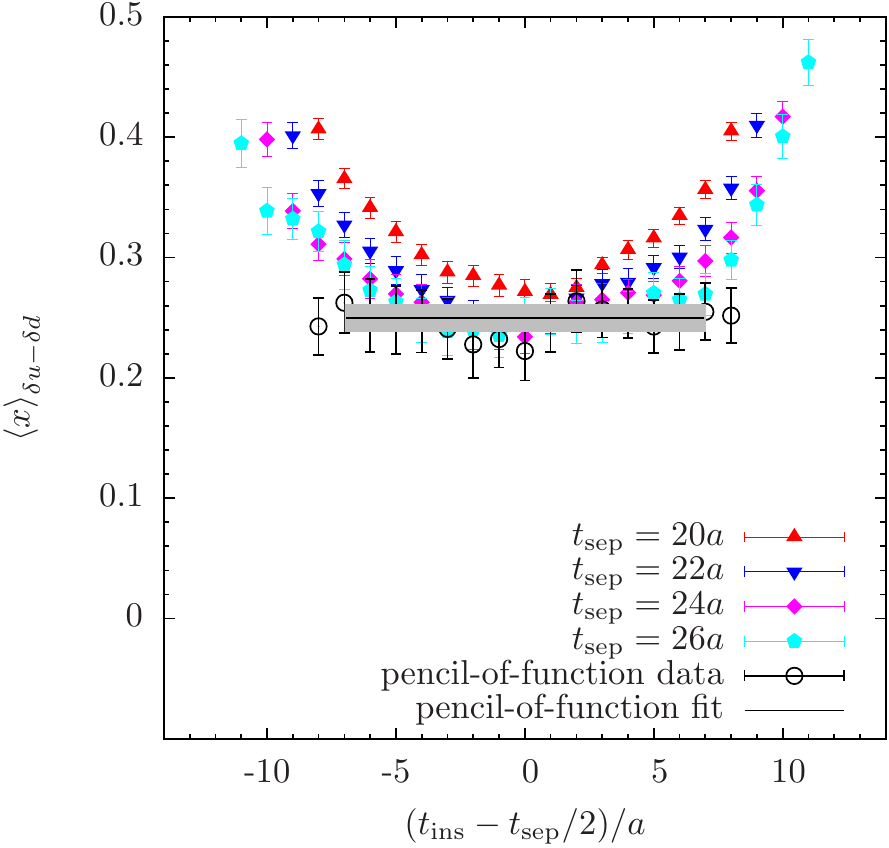}}\quad
 \subfigure[Results for $\avgx{-}{\delta}$ on J303 for different methods.]{\includegraphics[totalheight=0.275\textheight]{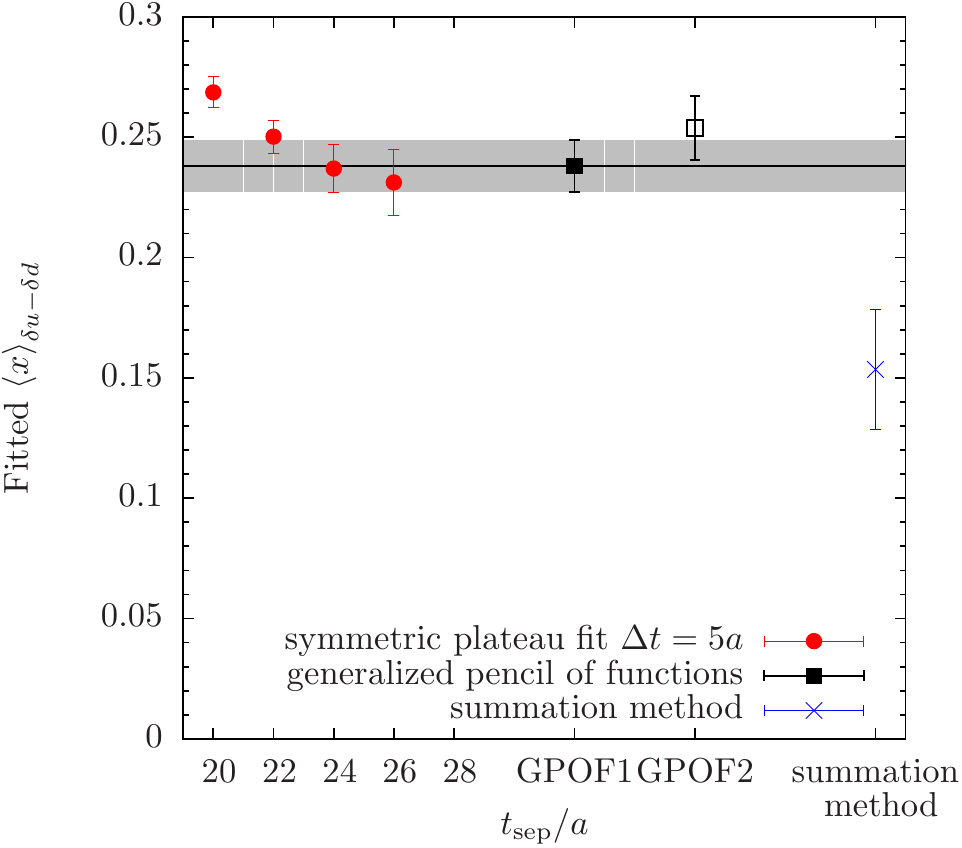}}
 \caption{Data for the ratios for $\avgx{-}{}$, $\avgx{-}{\Delta}$ and $\avgx{-}{\delta}$ on three ensembles and all values of $\tsep$ (left column). We also included the data and fit for the GPOF approach. The right column contains an overview of corresponding results from different methods. Results for the plateau method are plotted vs. $\tsep$. GPOF1 (GPOF2, open symbols) refers to using the smaller (larger) three values of $\tsep$, in the GPOF operator construction.}
 \label{fig:plateaux_and_methods}
\end{figure}

In the left column of Fig.~\ref{fig:plateaux_and_methods} we show results for the ratios in Eqs.~(\ref{eq:ratio_vD}),~(\ref{eq:ratio_aD})~and~(\ref{eq:ratio_tD}) at different values of $\tsep$ and for three different ensembles. We have also included the data for the corresponding ratios from the GPOF approach in Eq.~(\ref{eq:GPOF_ratio}), where the three lower values of $\tsep$ have been used in the operator construction. The amount of excited-state contamination is found to be rather similar for all three observables. Comparing results for different values of $M_\pi$ (e.g. for D200 and N200), we find that excited-state contamination is generally more severe at lighter pion masses, as expected. \par

In the right column of Fig.~\ref{fig:plateaux_and_methods} we compare results from different methods for the same observable and on the same ensemble. While there usually is a clear trend from the plateau method towards smaller results for increasing values of $\tsep$, it is not obvious that we can reach sufficient convergence to the ground state from this method. The summation method usually follows the apparent trend from the plateau method, leading to values that are sometimes significantly smaller than the results from the plateau method, e.g. for $\avgx{-}{}$ on D200. However, the errors are rather large and the results from the summation method are not very stable, e.g. leaving out the data at the lowest value of $\tsep$ can lead to very different results. Therefore, it is not clear if the available values of $\tsep$ are always in a regime where the summation method works sufficiently well. This issue will require further investigation, including more statistics over a wider range of source-sink separations. Interestingly, in some cases the GPOF approach favors values that are actually larger than the ones obtained from the plateau method at some of the larger values of $\tsep$. However, varying the operator basis for the GPOF approach with respect to the values of $\tsep$ that are used in the operator construction has rather little impact on the final results and does not reveal any particular trend. \par 

\clearpage
\begin{figure}[thb]
 \centering
 \subfigure[$\avgx{-}{}$ as function of $\Mpi^2$.]{\includegraphics[totalheight=0.275\textheight]{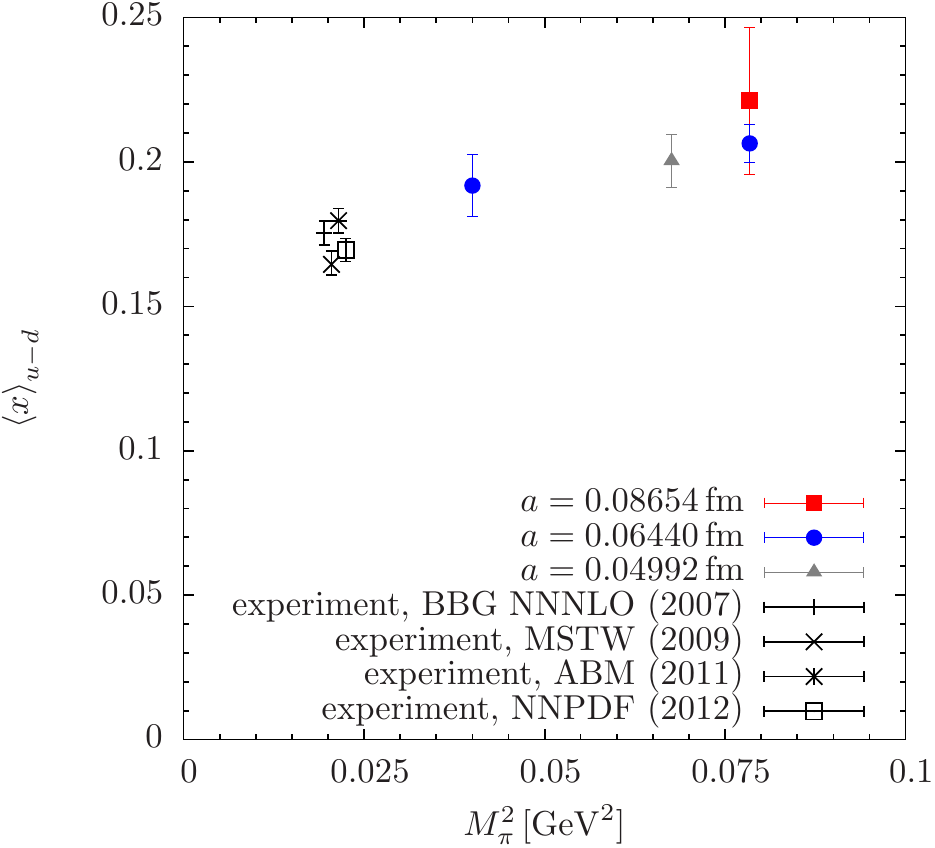}}
 \subfigure[Same as (a) but including data from other groups.]{\includegraphics[totalheight=0.275\textheight]{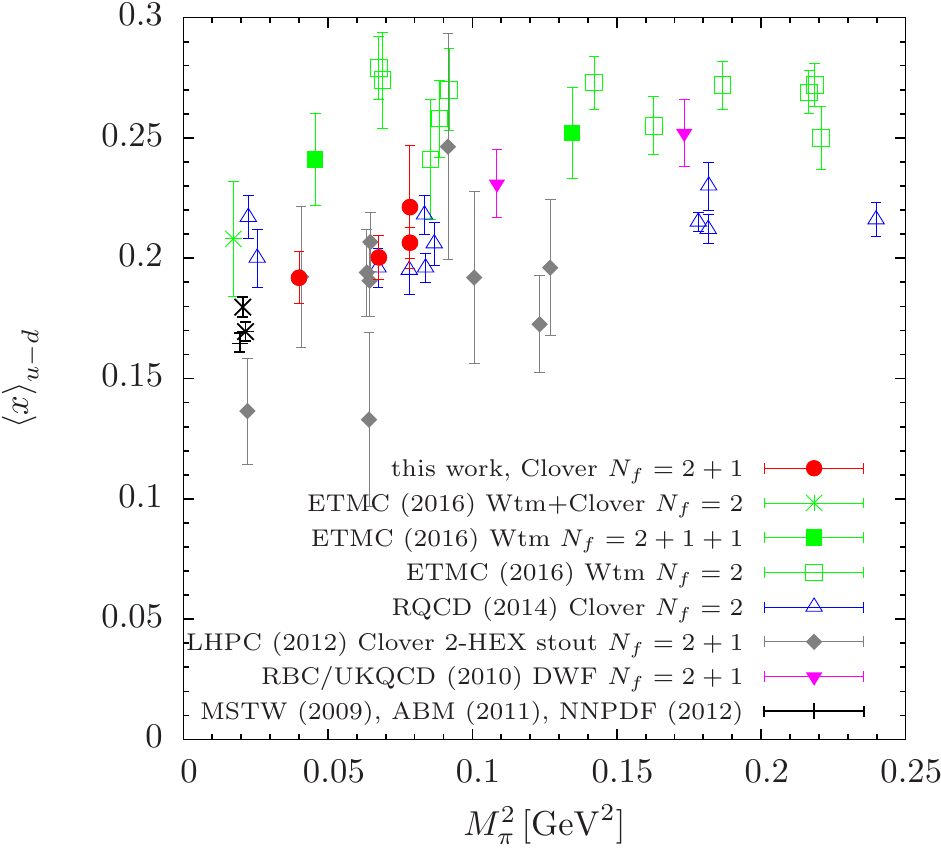}} \\
 \subfigure[$\avgx{-}{\Delta}$ as function of $\Mpi^2$.]{\includegraphics[totalheight=0.275\textheight]{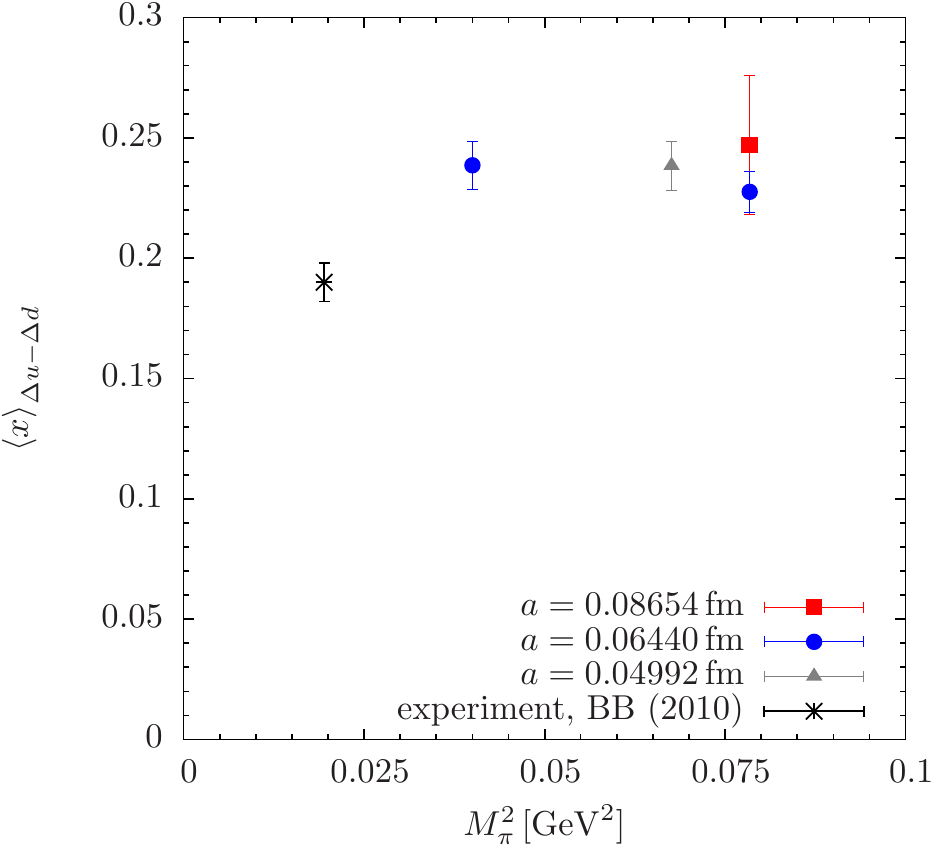}}
 \subfigure[$\avgx{-}{\delta}$ as function of $\Mpi^2$.]{\includegraphics[totalheight=0.275\textheight]{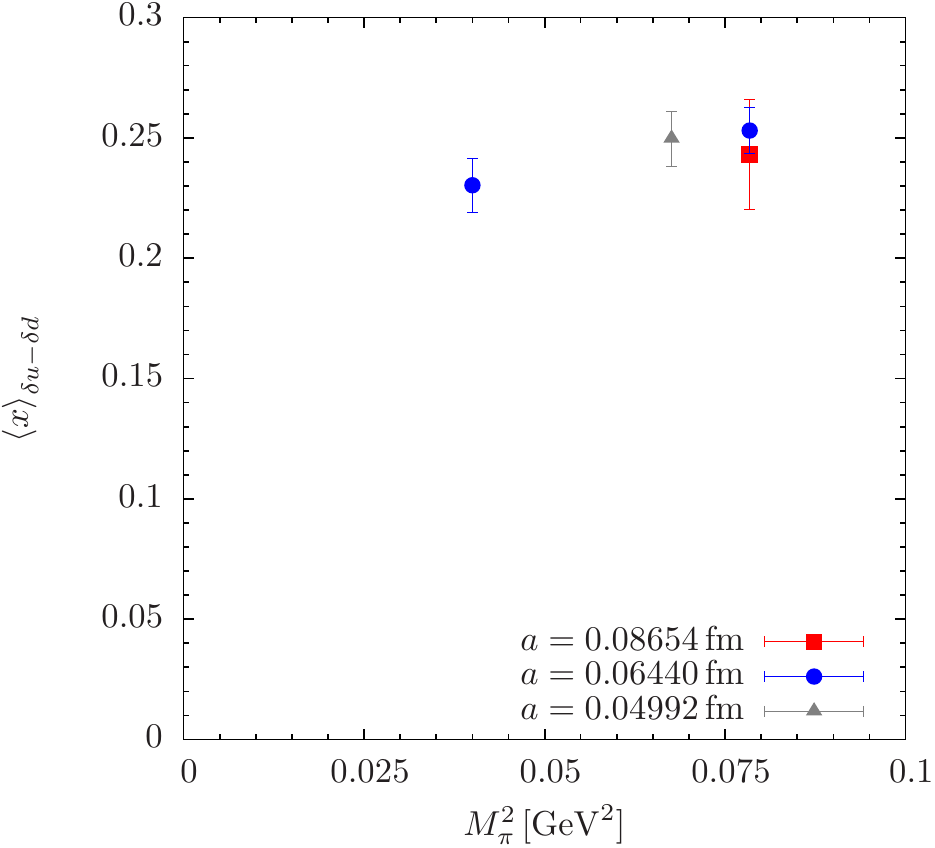}}
 \caption{Final results for $\avgx{-}{}$, $\avgx{-}{\Delta}$ and $\avgx{-}{\delta}$ from GPOF method as a function of $M_\pi^2$. Experimental data for $\avgx{-}{}$ and $\avgx{-}{\Delta}$ are taken from Refs.~\cite{Blumlein:2006be,Martin:2009bu,Alekhin:2012ig,Ball:2011uy} and Ref.~\cite{Blumlein:2010rn}, respectively. Additional lattice results shown in the upper right panel are taken from Refs.~\cite{Abdel-Rehim:2015owa,Bali:2014gha,Green,Aoki:2010xg}.}
 \label{fig:chiral_behavior}
\end{figure}
For now, we decided to use the GPOF approach with the lower three values of $\tsep$ to obtain preliminary results. In the upper left panel of Fig.~\ref{fig:chiral_behavior} we show the chiral behavior of our results. While we do not observe any dependence on the lattice spacing within errors, the behavior as a function of $M_\pi^2$ indicates rather good agreement with experimental results. In the upper right panel, we included recent lattice results from other collaborations. We find qualitative agreement with the results from RQCD, which employ Wilson Clover fermions with two dynamical quark flavors, as well as with the ones from LPHC. However, the ETMC results exhibit significantly larger values, although their value at the physical point is again compatible within its large error. Note that the LHPC results shown in this plot have been obtained from the summation method, which explains their relatively large errors \cite{Green}. In the lower two panels of Fig.~\ref{fig:chiral_behavior} we have plotted the chiral behavior of $\avgx{-}{\Delta}$ and $\avgx{-}{\delta}$, which are both experimentally less well known than $\avgx{-}{}$. \par

\clearpage
As mentioned before, we are currently adding more ensembles including a fourth lattice spacing. This will enable us to perform a reliable chiral and continuum extrapolation. Moreover, we will include more and larger source-sink separations to improve our control of excited-state effects. In particular, this will allow us to assess the stability and reliability of the summation method and the GPOF approach. For the latter it will also give more flexibility, e.g. for the choice of the time-shift $\Delta t$ in the operator construction. Besides, we are exploring two- or multi-state fits with the lowest energy gap fixed to its theoretically expected value. This will yield another cross-check and should ultimately lead to better control over systematics related to excited states. Finally, we intend to include quark-disconnected diagrams and extend our analysis to isoscalar operator insertions.

\section*{Acknowledgments}
This research is supported by the DFG through the SFB 1044. Calculations for this project were partly performed on the HPC clusters ``Clover'' at the Helmholtz-Institut Mainz and ``Mogon 2'' at JGU Mainz. Additional computer time has been allocated through projects HMZ21 and HMZ36 on the BlueGene supercomputer system ``JUQUEEN'' at NIC, J\"ulich. We are grateful to our colleagues in the CLS initiative for sharing ensembles.
\bibliography{lattice2017}

\end{document}